# Current-driven domain wall motion across a wide temperature range in a (Ga,Mn)(As,P) device

K. Y. Wang[1,2], K. W. Edmonds[3,1], A. C. Irvine[4], G. Tatara[5], E. De Ranieri[2,4], J. Wunderlich[2], K. Olejnik[2], A. W. Rushforth[3], R. P. Campion[5], D. A. Williams[2], C. T. Foxon[3] and B. L. Gallagher[3]

[1]*SKLSM, Institute of Semiconductors, CAS, P. O. Box 912, 100083, Beijing, P. R. China*

[2]*Hitachi Cambridge Laboratory, Cambridge CB3 0HE, United Kingdom*

[3] *School of Physics and Astronomy, University of Nottingham, Nottingham NG7 2RD, United Kingdom*

[4] *Microelectronics Research Centre, University of Cambridge, CB3 0HE, United Kingdom*

[5] *Graduate School of Science, Tokyo Metropolitan University, Hachioji, Tokyo 192-0397, Japan*

*Current-driven magnetic domain wall motion is demonstrated in the quaternary ferromagnetic semiconductor (Ga,Mn)(As,P) at temperatures well below the ferromagnetic transition temperature, with critical currents of the order $10^5 Acm^{-2}$. This is enabled by a much weaker domain wall pinning compared to (Ga,Mn)As layers grown on a strain-relaxed buffer layer. The critical current is shown to be comparable with theoretical predictions. The wide temperature range over which domain wall motion can be achieved indicates that this is a promising system for developing an improved understanding of spin-transfer torque in systems with strong spin-orbit interaction.*

The manipulation of magnetic domain walls (DWs) by spin transfer torque is one of the most active topics in the field of spintronics, driven by the prospect of potential ultrafast and high-density magnetic memory applications. While application-led research has mostly focused on current-driven DW motion in ferromagnetic transition-metal nanowires, studies of diluted ferromagnetic semiconductors such as (Ga,Mn)As offer the prospect of deep physical insights and routes to minimization of the critical current $J_c$ required for DW depinning. Experimentally, $J_c$ of around $10^5$ A/cm$^2$, two orders of magnitude lower than the typical value in metals, have been observed for (Ga,Mn)As nanowires [1,2,3]. Theoretical studies accounting for the strong spin-orbit interaction of $p$-type carriers in (Ga,Mn)As have pointed to a large $\beta$ or perpendicular spin-torque term which can substantially modify the critical current and DW velocity [4,5]. Experimental estimates of the $\beta$ term have varied from 0.01 or less to 0.36 [1,2,6,7].

The majority of experiments on (Ga,Mn)As structures have been performed on layers grown on a strain-relaxed (In,Ga)As buffer layer on GaAs(001), where the resulting tensile strain in the (Ga,Mn)As layer induces a perpendicular magnetic anisotropy. However, current-driven DW motion has only been observed at temperatures $T$ close to the ferromagnetic transition temperature $T_C$ (at $T>0.9T_C$ in typical cases [1,2,7] and $T>0.6T_C$ in a recent study [8]), where the magnetization is varying rapidly with $T$. This may be due to the high density of defects in the relaxed buffer layer, which thread through to the (Ga,Mn)As layer and act as pinning centers for DWs [2,9,10]. Furthermore, if the buffer layer relaxation and resulting tensile strain are anisotropic, this can lead to a large hard-axis anisotropy field which counteracts the spin-transfer torque [11]. The lack of available measurements of DW motion across a wide temperature range is an obstacle to the building of a detailed theoretical understanding of the observed low $J_c$.

Here, we investigate current-driven DW motion using an alternative diluted ferromagnetic semiconductor system, in which perpendicular magnetic anisotropy is achieved without requiring a strain-relaxed buffer layer. As reported recently, substitution of P on the group V sites of (Ga,Mn)As layers grown directly onto a GaAs(001) substrate can result in a large and tunable tensile strain giving rise to a perpendicular easy axis of magnetization [12,13]. Our key finding is that this results in a much lower DW pinning potential, enabling current-driven DW motion at temperatures much lower than $T_C$.

The 25nm (Ga,Mn)(As,P) film was grown by molecular beam epitaxy on a semi-insulating GaAs(001) substrate. The nominal Mn and P concentrations were estimated from the beam equivalent pressures to be 6% and 10% respectively. Details of the growth and magnetic properties of the layer are given elsewhere [12]. The sample was annealed in air for 24 hours at 190$^o$C in order to remove compensating Mn interstitial defects, and patterned into a Hall bar structure of width 4 $\mu m$, with voltage probes separated by 20 $\mu m$, and with the current orientation along [110] direction, using electron-beam lithography (fig. 1). A 10-15 $nm$ surface layer was etched away in rectangular regions situated on both ends of the Hall bar, in order to generate DW nucleation areas due to the reduced switching field in the etched regions. DWs can be pinned at the boundary between etched and un-etched area by applying an external magnetic field close to the nucleation field of the etched area. The current-driven DW motion was studied by using polar magneto-optical Kerr microscopy (PMOKM) and magnetotransport measurements. For comparison, we also present results for a (Ga,Mn)As/(In,Ga)As/GaAs(001) Hall bar device with the same structure and same nominal Mn concentration. A detailed investigation of the current-driven DW motion in this control device has been presented elsewhere [3]. The $T_C$ of the (Ga,Mn)As and (Ga,Mn)(As,P) devices are 122K and 92K, respectively, as determined by PMOKM.

The critical current $J_c$ required to de-pin a DW from the boundary between the etched and unetched regions is plotted as a function of temperature for the studied devices in Fig. 2(a). This was measured in the following way. Firstly, a magnetic field, applied along the perpendicular easy magnetic axis, is swept up to the opposite switching field of the etched region and back down to zero in order to initialize a DW at the boundary region. Then, a dc electric current is ramped up from zero at a rate of $2\times10^3$ $A$cm$^{-2}$s$^{-1}$, while monitoring the voltage at the Hall contacts closest to the etched region. As soon as an abrupt change in the Hall voltage is detected, the electric current is reduced to zero, and the magnetic configuration of the device is imaged using PMOKM before and after the dc current. The light is screened while the current is applied in order to reduce the optical excited effects. The sample temperature during the application of current is corrected to account for Joule heating, using the longitudinal resistance of the device as described elsewhere [3].

As shown by Fig. 2(a), current-driven domain wall motion is observed in the (Ga,Mn)(As,P) device at temperatures down to 25K ($T/T_C$=0.28, which is the lowest temperature accessible in our system, after accounting for Joule heating), with critical currents below $8\times10^5$Acm$^{-2}$. This is further demonstrated by Fig. 1, which shows PMOKM images of the (Ga,Mn)(As,P) device before and after application of the current at a temperature of 25.8K. Consistent with previous studies of (Ga,Mn)As devices [1,3], the DW always moves in the opposite direction to the applied current. For the (Ga,Mn)As device, $J_c$ rises sharply as the temperature is decreased, and current-driven DW motion cannot be observed for temperatures below 100K ($T/T_C$=0.8).

In order to determine whether the differences in the DW motion for the two samples are due to a stronger extrinsic DW pinning in the layer grown on the strain-relaxed buffer layer, we measured the pinning potential for a DW trapped close to the boundary region. First, the DW configuration is initialized as before, and then the external magnetic field is steadily increased until it reaches the depinning field $H_{de}$ where the DW starts to move away from its initial position. The pinning potential is defined as $V_0=M*H_{de}$, where $M$ is the magnetization. Figure 2(b), (c) and (d) show the temperature dependence of the magnetization (measured by PMOKM and calibrated to the low temperature value obtained by SQUID magnetometry), $H_{de}$ and $V_0$ respectively. It can be seen that the depinning field and pinning potential rise rapidly with decreasing temperature in the (Ga,Mn)As device, and are an order of magnitude larger than in the (Ga,Mn)(As,P) device even at temperatures close to $T_C$. This indicates that it is the large pinning potentials due to the high defect density in the (Ga,Mn)As/(In,Ga)As structure that prevents efficient current-induced DW motion. Note that above ~$0.9T_C$, it is not possible to form a stable domain wall at the boundary between etched and unetched regions in the (Ga,Mn)(As,P), due to the very weak pinning.

Tatara and Kohno [11] obtained an analytic expression for the intrinsic critical current in a ferromagnetic nanowire, in the case that the pinning potential is weak. For (Ga,Mn)(As,P) with zinc-blende structure this is given by

$$J_{ic} = eS K_\perp \lambda / \hbar P$$

where $S$ is the spin magnitude per Ga site, $P$ is the current spin polarization, $K_\perp$ is the hard axis magnetic anisotropy energy density, and $\lambda$ is the domain wall width. $K_\perp$ is given by ½$MH_A$, where $H_A$ is the magnetic field required to rotate the magnetization in the plane of the Hall bar device. The procedure assumes that the x-axis is the in-plane magnetic easy axis. $H_A$ is estimated from anisotropic magnetoresistance (AMR) measurements, shown in Fig. 3, for field sweeps along three orthogonal directions. At high field, the slopes of the three magnetoresistance curves are similar. The anisotropic magnetoresistance manifests itself at low fields, where the magnetization rotates away from the easy magnetization direction and into the direction of the magnetic field. The magnetization is fully aligned along a hard axis direction at a field $H_{sat}$ where the slope of the magnetoresistance matches that obtained for the perpendicular easy axis. We obtain $H_A$ from the difference in $H_{sat}$ between the two in-plane directions ($x$ and $y$ in Fig. 3). At $T$=30K, we find $H_A$=400±100 Oe for the (Ga,Mn)(As,P) device. $K_\perp$ obtained here is in agreement with the value obtained by using the method in Ref. 1.

In order to estimate the expected size of $J_{ic}$, we assume that $S$ and $P$ track the bulk magnetization ($S=S_0 M/M_0$ and $P=P_0 M/M_0$), with $S_0 \approx 0.2$ and $P_0$ reported to be around 0.8 from spin-injection measurements [14]. The domain wall width is predicted to be of order 10nm for (Ga,Mn)(As,P) materials with perpendicular magnetic anisotropy [15].

Combined with the measured $H_A$, this leads to an estimated $J_{ic}$ of $2.5 \times 10^5$ Acm$^{-2}$ at 30K, in reasonable agreement with the experimental result given the uncertainties involved.

In summary, we have demonstrated current-driven domain wall motion across a wide range of temperatures in a (Ga,Mn)(As,P) device. Studies of this material should lead to an improved understanding of spin transfer torque in ferromagnetic semiconductor nanowires, including the influence of thermal fluctuations, adiabatic versus non-adiabatic spin torques, and the role of the spin-orbit interaction.

This project was supported by EU Grants IST-015728 and 214499, EPSRC-UK (EP/H002294/01) and EPSRC-NSFC joint grant 10911130232/A0402. KYW acknowledges support of Chinese Academy of Sciences "100 talent program". KWE acknowledges support of Chinese Academy of Sciences Visiting Professorship for Senior Foreign Scientists. GT acknowledges support from Grant-in-Aid for Scientific Research in Priority Areas (1948027), the Kurata Memorial Hitachi Science and Technology Foundation and the Sumitomo Foundation. AWR acknowledges support of an EPSRC Career Acceleration Fellowship (EP/H003487/1).

**Figure captions**

FIG. 1. (Color online) PMOKM images of the (Ga,Mn)(As,P) Hall bar device before (a) and after (b) application of the critical current at $T$=25.8K. The arrow marks the direction of the applied current. The dashed lines mark the positions of the etching step used for domain nucleation.

FIG. 2. (Color online) Dependence on reduced temperature $T/T_C$ of (a) critical current $J_c$, (b) magnetization $M$, (c) depinning field $H_d$, and (d) pinning potential $V_0=MH_d$. Filled circles are for the (Ga,Mn)(As,P) device ($T_C$=92K) and unfilled circles are for the (Ga,Mn)As device ($T_C$=122K).

FIG. 3. (Color online) Magnetoresistance of the (Ga,Mn)(As,P) device at 30K for field sweeps along three orthogonal axes ($x$, $y$ and $z$, illustrated in the inset), where the arrows indicate the $H_{sat}$ and the scanned field from positive to negative is shown in black curves.

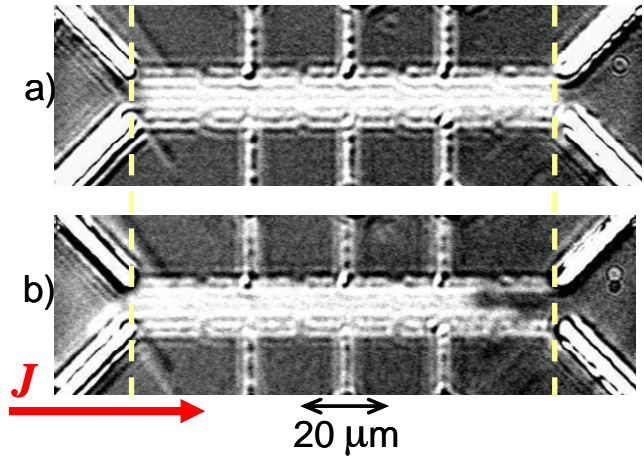

*Figure 1. K. Y. Wang et al.*

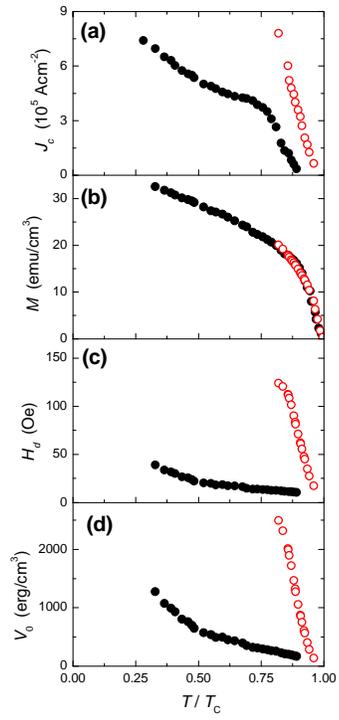

Figure 2. K. Y. Wang et al.

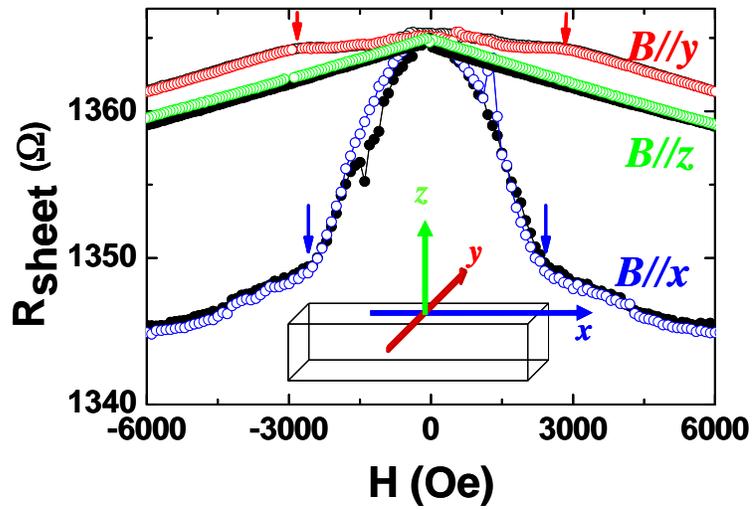

*Figure 3 K. Y. Wang et al.*